\documentstyle[preprint,eqsecnum,aps]{revtex}
\begin{document}
\draft
%\tighten
\preprint{UCSBTH-96-24, hep-th/9610171}
\title{Where is the Information Stored in Black Holes?}
\author{Gary T. Horowitz\cite{gary}} 
\address{Physics Department, University of California,
Santa Barbara, California 93106} 
\author{Donald Marolf\cite{don}}
\address{Physics Department, Syracuse University, Syracuse, New York 13244}
%\date
\maketitle

\begin{abstract}
It is shown that many modes of the gravitational
field exist only inside the horizon of an extreme black hole
in string theory.  At least  in certain cases,
the number of such modes is sufficient to account for the Bekenstein-Hawking
entropy.  These modes are associated with sources which carry
Ramond-Ramond charge, and so may be viewed as the strong coupling limit
of D-branes.
Although these sources naturally
live at the singularity, they are well defined and generate modes
which extend out to the horizon. This suggests that the information
in an extreme black hole 
is not localized near the singularity or the horizon,
but extends between them.

\end{abstract}

\newcommand{\s}{\sigma}
\newcommand{\p}{\partial}
\newcommand{\pp}{\partial_+}
\newcommand{\pk}{\partial_-}
\newcommand{\V}{{\sf V}}
\newcommand{\HH}{{\cal H}}
\newcommand{\be}{\begin{equation}}
\newcommand{\ee}{\end{equation}}
\newcommand{\lp}{\left (}
\newcommand{\rp}{\right )}

\vfil
\eject

\baselineskip = 16pt
\section{Introduction}

A key ingredient in understanding the  black hole information puzzle \cite{pre}
is the question of where the states accounting for the 
black hole entropy are located.  If the Bekenstein-Hawking entropy 
\cite{bek,haw}
is associated with the matter that forms the black hole,
then it would appear that these states are
localized near the singularity. On the other hand, since the entropy
is proportional to the horizon area, it has been suggested that these
states are associated with horizon fluctuations \cite{zuth}.
The resolution is important for deciding whether information is lost
in black hole evaporation, since storing the information near
the singularity would make it difficult to be recovered
without violating causality or locality. 
This, of course,
would not be such a problem if information were stored near the horizon.

Recently, the states associated with extremal and near extremal black holes
have been identified in weakly coupled string theory \cite{stva,cama,host,hor}.
The number of such
states exactly reproduces the black hole entropy (for large black holes).
While there are plausible arguments for extrapolating the {\it number} of
states from weak coupling to strong coupling, there has so far been
little discussion of what  these weak coupling states look like at strong
coupling, where a large black hole is present. 

One can find arguments which support either possibility, that these
states are localized near the singularity or near the horizon. 
For example, at weak coupling,
one considers bound states of D-branes \cite{pol,pcj}
which carry the same charges 
as the black hole. Although the size of such bound states is not known,
it is expected to be no larger than the string length, which is 
set by the string
tension. At strong coupling, the event horizon
is much larger than the string length. Since the D-branes carry
the charge, and the source of the charge in the black hole is the
singularity, the D-branes must lie at the singularity. Thus the
bound states of D-branes should be localized to within a string length of
the singularity.
However, the validity of the laws of black hole thermodynamics
suggest that the Bekenstein-Hawking entropy should be associated 
with states that are accessible to (i.e. can interact with) external
probes \cite{rov}. This seems to imply that the states are localized near the
horizon.

We will argue that there is another possibility which combines
the desired features of both alternatives. We will show that 
the gravitational field has a large number of modes
which exist only inside the horizon of
an extreme black hole. Since there is a timelike singularity inside the
horizon, it is perhaps not surprising that additional
modes exist. What {\it is} surprising is that the modes we consider
do not propagate into
another asymptotically flat region of spacetime, but are entirely contained
within the horizon. The radial profiles of these interior modes are fixed and
they extend from the singularity to the horizon. We will also show that
there is a well defined sense in which 
these modes are generated by sources living at the
black hole singularity. Since these sources carry Ramond-Ramond charge,
it is natural to interpret them as the strong coupling limit of D-branes.
Thus although the
D-branes live at the singularity, they couple to modes which extend out
to the horizon and carry the information about their state of excitation.

The interior modes we consider differ  from waves outside
the horizon \cite{lawi,lawiII,homa,homaII} in several respects.
The exterior waves are all homogeneous
in the compactified dimensions of the spacetime, while the interior waves 
are not. There are thus many more waves which exist inside the horizon.
Although we cannot do a precise counting at this time, it is clear
that the number of these modes is sufficient to account for the
Bekenstein-Hawking entropy (at least when the black hole carries
unit fivebrane charge).
More importantly, it has recently
been shown that, although the metric for the exterior waves  is continuous
on the horizon, there is a mild curvature singularity there\footnote{Despite
this singularity, the area of the horizon is well defined and agrees with
the counting of D-brane states \cite{homa,homaII}.
The physical significance of this singularity
is being investigated.} \cite{hoya}. The
interior modes we discuss will leave the metric at least $C^2$ on the
horizon, so that no curvature singularity is generated.

The outline of this paper is as follows. We begin
in section II by reviewing the exterior solutions of the extreme black holes
we wish to consider, and describe the simplest (homogeneous)
interior solutions.  In section III, we present the equations governing
a general class of interior modes and
show that these modes can be associated with sources that live
on the singularity. In the next section we discuss stationary
solutions of these equations, and the smoothness of the horizon.
In section V we add time dependence and discuss the connection with
black hole entropy. In particular, the states associated with  oscillating
D-branes are described. Some concluding remarks and open questions are
contained in section VI. The details of the matching conditions at
the horizon are contained in appendix A. As a first step toward understanding
how these interior modes might be excited,
we study  an oscillating test string as it falls into the
black hole in appendix B.

\section{Homogeneous Interiors}

We begin our investigation of the region behind the event horizon by
considering the simplest 
homogeneous cases.  Such solutions may be obtained, for example, 
by analytic continuation of an exterior solution through the horizon.
This of course requires the use of coordinates in which the
metric is in fact analytic at the horizon.

Let us therefore begin by reviewing the exterior solutions. 
The low energy action for the type IIB string theory contains
the terms (in the Einstein frame)
\be \label{action}
S = {1\over 16\pi G} \int d^{10} x \sqrt{-g}\left(
R -  {1\over 2} (\nabla \phi)^2
- {1\over 12}  e^\phi\ {\cal H}^2 \right )
\ee
where $\phi$ is the dilaton and ${\cal H}$ is the Ramond-Ramond three form.
We will consider solutions on $T^5 \times {\bf R}^5$, 
which reduce to black holes in $4+1$ dimensions.
Following the conventions of our earlier papers \cite{homa,homaII}, 
we label the 4+1 asymptotically flat
`external' directions by coordinates $(x^i,t)$, and divide the five
torus into an $S^1$ (labeled by the coordinate $z$) and a $T^4$ (labeled
by the coordinates $y^i$).  We take the $S^1$ to have coordinate
length $L$ and the $T^4$ to have coordinate volume ${\sf V}$.
It is often useful to think of the 
solution as corresponding to a black {\it string} in $5+1$ dimensions, 
where the string lies along the $z$ axis.  In such a picture the
$T^4$ would be considered an `internal' four torus.  These solutions
carry electric and magnetic charge with respect to the 
three form ${\cal H}$. They also carry momentum in the $z$ direction.
At weak coupling, these  charges are reproduced 
by D-onebrane and D-fivebrane
sources, with the fivebranes lying in the $(t,z,y^i)$ 5+1 space
and the onebranes lying in the $(t,z)$ plane. 

Introducing the coordinates
$u=t-z$ and $v = t+z$, the exterior black hole
solution takes the form \cite{tse,cama}

\begin{equation}
\label{outside}
ds^2 =  H_1^{1/4} H_5^{3/4} \left[ {{du} \over {H_1H_5}} \left(-dv
+ K du \right) + {{dy_i dy^i} \over {H_5}} + dx_i dx^i \right]
\ee
\be
e^{-2\phi} = {H_5 \over H_1}
\ee
\be
\label{Heqn}
 \HH_{auv} = H_1^{-2}\p_a H_1, \qquad \HH_{ijk} = -\epsilon_{ijkl}\p^l H_5
 \ee
where $\epsilon_{ijkl}$ is the flat space volume form on the $x$ space
 (the indices $i,j,k,l$ in (\ref{Heqn}) refer to the $x$ space) and
 $a$ runs over all $x^i,y^i$.
The functions $H_1$, $H_5$, and $K$ are given by
\be
 H_1 = 1 + {{r_1^2} \over {r^2}}, \qquad
H_5 = 1 + {{r_5^2} \over {r^2}}, \qquad K = {p\over r^2}
\ee
where  $r^2 = x_i x^i$, and the constants $r_1$, $r_5$, and $p$
determine the 
electric and magnetic charges of the three form, and the momentum respectively.
In this coordinate system, the horizon lies
at the coordinate singularity $r=0$. Its area is given by $A = 2 \pi^2
r_1 r_5 {\sf V} L \sqrt{p}$. The nonextremal black hole solutions are also
known \cite{cvyo,hms} but will not be considered here.

New coordinates may be introduced such that
the metric is analytic at the horizon (see e.g. the appendix of \cite{homa}).
It can then be continued into the
interior.  A further change of coordinates places the interior metric
in the convenient form:

\begin{equation}
\label{inside}
ds^2 =  H_1^{1/4} H_5^{3/4} \left[ {{du} \over {H_1H_5}} \left(dv
+ {p \over {r^2}}du \right) + {{dy_i dy^i} \over {H_5}} + dx_i dx^i \right]
\end{equation}
with {\it exactly} the same coordinate identifications (corresponding
to the compact directions) as in (\ref{outside}).  However, we now
have
\be\label{minside}
H_1 = {{r_1^2} \over {r^2}} -1, \qquad H_5 = {{r_5^2} \over {r^2}}-1.
\ee
Note that, aside from the form of the harmonic functions $H_1$ and
$H_5$, the only difference between (\ref{inside}) and (\ref{outside})
is a single negative sign in the $dudv$ term.
As a check on these formulas, note that for the case $r_1 = r_5 \equiv r_0$,
both metrics (\ref{inside}) and (\ref{outside}) reduce to the more
familiar form of the solution
\be\label{both}
    ds^2 = - \lp 1-{r_0^2\over \hat r^2}\rp dudv +{p\over \hat r^2} du^2
+ \lp 1-{r_0^2\over \hat r^2}\rp^{-2} d\hat r^2 + \hat r^2 d\Omega_3^2
	  +dy_i dy^i
\ee
where $r^2 = \hat r^2 - r_0^2$ in the exterior and $r^2 = r_0^2 -\hat r^2$
in the interior.  

In the interior metric (\ref{inside}) the horizon again lies at $r=0$.  
Thus, this coordinate 
system is `inside-out' in the sense that moving to larger values of $r$
corresponds to moving deeper into the interior.  Note that the
singularity lies at $r=r_1$ or $r=r_5$ (whichever is smaller) 
where $H_1$ or $H_5$ vanishes.
This is a timelike curvature singularity, much
like the singularity of usual extreme Reissner-Nordstr\"om black holes.
As a result of the change of sign in the $dudv$ term,
the coefficient of $dz^2$ is proportional to $({p \over {r^2}} -1)$
which becomes negative for $r > \sqrt{p}$.  Thus, if $\sqrt{p} < r_1,r_5$ the
physical region $0 < r < r_1,r_5$ contains closed timelike curves.

\section{More General Interior Solutions and Sources}

Although the interior solution described above is the unique analytic
extension of the exterior solution, it turns out that there are many
other interior solutions which leave the horizon nonsingular. We now
discuss these more general solutions and show that they can be viewed
 as arising from sources at the singularity.
Since these sources carry RR charge, they are naturally interpreted as
the strong
coupling limit of D-branes.

We will consider solutions of the form
\begin{equation}
\label{ansatz}
ds^2 = H_1^{1/4} H_5^{3/4} \left[   {{du}
\over {H_1 H_5}}(\epsilon dv  + K du  + 2A_i dy^i) + {{dy_i dy^i} \over {H_5}}
+ dx_i dx^i   \right] 
\end{equation}
\be
e^{-2\phi} = {H_5 \over H_1}
\ee
\be\label{ansatz2}
 \HH_{auv} = H_1^{-2}\p_a H_1, \qquad  \HH_{aub} = 2 \p_{[a} A_{b]},
 \qquad \HH_{ijk} = -\epsilon_{ijkl}\p^l H_5
 \ee
where $H_1 = H_1(u,x,y)$, $H_5 = H_5(x)$, $K= K(u,x,y)$, $A_{y^i} = A_{y^i}
(u,x,y)$, and the indices $a,b$ run over $x^i, y^j$ though
$A_{x^i} = 0$.
This is a generalization of solutions that have been considered 
previously \cite{gar,bko,hots,cmp,cvts}.
This ansatz preserves the null translational symmetry $\p/\p v$ of
the original solution. 
Here $\epsilon$ is a sign which is clearly arbitrary
(as it may be changed by sending $v \rightarrow -v$), but which must be
reversed (as in section II) when
matching interior and exterior solutions.  The motivation for this
form of the solution comes from the description of the microstates of
this black hole in the limit of weak string coupling.
For a single fivebrane,
the entropy comes from the oscillations of onebranes inside the fivebrane.
Since $H_5$ is associated with the fivebrane (this will be made more precise
below) which does not oscillate, we have assumed that it is only a function
of $x$. The vector $A_i$ describes momentum flow in the $i^{th}$ direction.
Since the onebranes only oscillate inside the fivebrane, we have assumed
that $A_i$ has components only in the internal ($y$) directions.

It may be verified that (\ref{ansatz})-(\ref{ansatz2}) is an
extremum of the action (\ref{action})
when the following five conditions are satisfied:
\begin{eqnarray}
\label{main}
\partial^2_x H_5 = 0  \cr 
\partial_x^2 H_1 + H_5 \partial^2_y H_1 = 0 \cr
\partial^2_x K + H_5 \partial_y^2 K + 2 H_5 \partial_u [ \partial_u
H_1 - \partial_i A^i ] = 0 \cr
\partial_{x^j} [ \partial_u H_1 - \partial_i A^i ] = 0 \cr
\partial^2_x A_i + H_5 \partial_y^2 A_i + H_5 
\partial_i [ \partial_u H_1 -
 \partial_j A^j ]
= 0
\end{eqnarray}
where the indices $i,j$
run over the four $y$ coordinates, $\partial^2_x$, $\partial^2_y$
are the {\it flat-space} Laplacians associated with the $x$ and $y$
coordinates, and the indices $i$ are raised and lowered
using the flat space Euclidean metric.
In particular, when $\partial_u H_1 = \partial_i A^i$ and the solutions are
independent of $y$, we find that $H_1$, $H_5$, $K$, and $A_i$ are just
flat-space harmonic functions of $x$.   Given that the null symmetry
$\p/\p v$ is preserved, it is likely that these solutions are supersymmetric.

Since we can solve for $H_5$ first, the above equations are all essentially
linear. As a result,
we may think of these  fields as produced by a set of
localized sources which lie at the singularity. Although the singularity
is pointlike in the physical spacetime metric, it  corresponds to
the vanishing of $H_1$ or $H_5$, which often occurs on
a surface of finite coordinate
size in the four dimensional Euclidean space parameterized
by $x$. In electrostatics, if one wants to solve Poisson's equation
with  a shell of
charge, one usually demands regularity at the origin,
so that the solution is trivial
inside and nontrivial outside the shell.  Here, the requirement of
an event horizon requires that the solution be nontrivial inside the
shell, so we can take it to be trivial outside.
For example, consider the homogeneous interior solution (\ref{inside}),
(\ref{minside})
with $r_1 < r_5$, so the singularity is at $r=r_1$. Away from the
singularity, $r<r_1$, this solution is identical to the solution $A_i =0$,
$$H_1 = \left(  {{r_1^2} \over {r^2}} - 1\right) \theta(r_1-r),
\ \  H_5 =  \left({{r_5^2} \over {r^2}} -1 \right) 
\theta(r_1 -r) +
\left( {{r_5^2} \over {r_1^2}} -1 \right)  \theta(r -r_1) $$
\be\label{srce}
 K = \left( {p \over {r^2}} \right) \theta(r_1-r) + \left(
{p \over {r_1^2}} \right) \theta(r-r_1)
\ee
where $\theta(r)$ is the usual step function.
The only difference is that the solution (\ref{srce}) has sources
at $r=r_1$ with just the right strength
to account for the total charges and momentum of the black string.  
Since the source
of the charges for the black string should lie at the singularity, (\ref{srce})
is a more accurate description of the physics. As the sources carry
RR charge, they can be viewed as the strong coupling limit of D-branes.

One might ask what would happen if one tried to place the sources
away from the singularity at some $r=r_0 < r_1$. This turns out to be unphysical
since the sources would have negative local energy density. This can be
seen as follows. If the sources are at $r=r_0 < r_1$, then 
the spacetime is nonsingular and 
$H_1$, $H_5$, and $K$ are all positive {\it constants} for $r> r_0$.
Hence,  this region of spacetime is completely flat. One thus obtains
a nonsingular solution with zero total energy and an event horizon.
The positive energy
theorem for black holes \cite{ghhp} states that any such spacetime 
must contain matter with negative local energy density.
The reason for the negative energy density can be understood physically
by considering the four dimensional example of a shell of $q=m$ dust.
The spacetime is the extreme Reissner-Nordstr\"om solution with charge $Q$
outside the shell and flat
space inside. When the shell is large, the energy in the electromagnetic
field is small and most of the energy comes from the shell. As we decrease
the radius of the shell, the total charge (and hence total mass) remains
constant, but the energy in the electromagnetic field increases. Thus the
energy density of the shell must decrease.  For small enough shells (inside
the horizon of the extreme black hole) the energy density must become 
negative.  This may be studied in detail by using the spherically
symmetric mass function appropriate to this spacetime. 

This result has implications for which D-branes can be placed in
static equilibrium around an extreme (positively charged) black hole.
Outside the horizon, there exist static, BPS solutions
with D-branes which have positive charge 
and energy density.
There are also static solutions with sources that have negative charge
and negative energy density, but these are not usually considered because they
are unphysical. Inside the horizon, the situation is reversed.
It is the negatively charged  D-branes which remain static (and have positive
energy density). A positively charged source could remain static inside the
horizon only if
it had negative energy density.

The idea that a  given set of fields ($H_1$, $H_5$, $K$, and $A_i$)
may be thought of as being produced by sources at the singularity in fact
holds much more generally. The trick is to choose the 
proper boundary conditions as,
in general, the solutions to (\ref{main})  will not all be constant at
the singularity, so they cannot be extended as constants beyond the
singularity.  Let us assume
$\p_u H_1 = \p_i A^i$.
We will see later that these are the solutions of greatest interest.
Then $H_1$, $K$, and $A_i$ are all determined by the same differential
operator
\be\label{deftil}
\tilde{\nabla}^2 = \partial_x^2 
+ H_5 \partial_y^2. 
\ee
The field
between the horizon and the singularity can be associated
with a unique set of sources at the singularity (and a unique
set of fields in the unphysical region) provided that the
equations 
\begin{eqnarray}\label{laplace}
\tilde{\nabla}^2 H_1 &=& 0 \cr
\tilde{\nabla}^2 K &=& 0 \cr
\tilde{\nabla}^2 A_i &=& 0 
\end{eqnarray}
are imposed everywhere except on the surface where the charge will lie
(the surface where $H_1$ or $H_5$ first vanishes) and provided that
the fields are required to approach constant values at large
$r$ and to carry no net (monopole) charge; i.e.
$H_1 \sim const + {\cal O}(r^{-3})$.

Two comments about this association between fields and sources
are now in order.  The first is that, in general, the above procedure 
may not preserve the condition $\p_u H_1 = \p_i A^i$  in the unphysical
region beyond the singularity. As a result, they are
not solutions
to the full equations of motion in this region. However, there is no
reason why the full equations of motion must be satisfied in the 
region beyond the singularity. If one wanted to keep the equations
satisfied, there are two possibilities. 
Since $A_i$
describes the momentum of the sources, roughly speaking the problem
is that some interior solutions describe sources which remain on some 
surface $S$ but have momentum transverse to this surface.  
This may be remedied by allowing
the sources of the $A_i$ fields to extend off of the singular surface, 
deeper into the unphysical region.  A simple counting of degrees of
freedom suggests that it can also be
remedied by generalizing our solutions to include
an $A$ field that also points in the $x$ directions (allowing oscillations
in the $x$ directions) and by considering solutions with
$\partial_u H_1 \not= \partial_iA^i $.
However, for the purposes
of the present work it will be sufficient to restrict attention to solutions
for which the above procedure does in fact generate fields which satisfy
$\partial_u H_1 = \partial_iA^i $, even in the unphysical
region.  This is because the most interesting solutions will in fact
have this property.

The other comment concerns the boundary conditions imposed
at large $r$.
Since this condition refers to the region beyond the singularity, it
is somewhat arbitrary.  However,
the fact that we must introduce
a boundary condition by hand is not unexpected -- it is simply due to
the presence of the timelike singularity.
We require the fields to approach constants
as this is a 
familiar boundary condition for elliptic equations and will
generate familiar relationships between the associated charge
densities and fields.  
The zero net charge restriction says that
the total charge found inside the horizon (i.e., in the $r >0$
region) is equal to the total charge registered on the horizon.
Finally, note that the interpretation of 
the fields as arising from sources at the singularity holds for a wide class
of boundary conditions.  It is only the exact form of the pairing 
between fields and sources that depends on the
particular boundary condition chosen.

\section{Inhomogeneous Interiors}

In this section we investigate interior solutions which are stationary,
i.e. $u$ independent, but inhomogeneous in the $y$ directions. The
effects of allowing $u$ dependence will be considered in the next section.
We will set $\partial_i A^i =0$, so the equations  (\ref{main}) decouple.
We also assume that all fields remain spherically symmetric in the four
dimensional $x$ space. This implies that $H_5$ keeps the same form that it had
in the homogeneous case: $H_5 = {{r_5^2 } \over {r^2}} - 1$.

The functions $H_1$, $K$, and $A_i$ then satisfy (\ref{laplace}).
Let us examine the behavior of the solutions near
the horizon of the black hole, $r=0$.  The coefficient $C_k(r)$ of the
Fourier mode $e^{ik_jy^j}$  satisfies
\begin{equation}\label{inhor}
r^{-3} \partial_r\lp r^3 \partial_r C_k\rp  =  k^2 {{r_5^2} \over 
{r^2}} C_k
\end{equation}
so that $C_k \sim r^\alpha$ where $\alpha = -1 \pm \sqrt{1 + k^2 r_5^2}$.
The exact solutions can be expressed in terms of
Bessel functions, although the detailed form of the solution will not
be important here.

For $k^2 \not= 0$, the modes
with $\alpha = -1 - \sqrt{1 + k^2 r^2_5} < -2$ cause the horizon to become
singular.  If
such modes are present in the $H_1$ field, then since
$e^\phi = \sqrt{H_1/H_5}$, such solutions have divergent 
dilaton on the horizon.  If such modes are present in $K$, then the
norm of the Killing field $\p/\p u$ diverges on the horizon. If they
are present in $A_i$ then, at
the very least, the structure of the horizon is radically altered from
the original homogeneous case.
It seems likely that the horizon would again become singular.
On the other hand, the modes with
$\alpha = -1 + \sqrt{1 + k^2 r_5^2}$ vanish at the horizon.  As
a result, they do not alter the properties of the horizon, at least
to leading order.  In particular, the horizon itself remains
homogeneous.   A more careful study of such modes is performed 
in appendix A where it is found that 
a mode which behaves as $e^{iky}$
(and thus has $\alpha = -1 + \sqrt{1 + k^2 r_5^2}$) is at
least $C^{\epsilon/2}$ for all $\epsilon < \alpha/2 -1$ at the horizon. 
Thus, 
if any mode with $\alpha > 6$
is added to the interior solution, the horizon remains at least
a $C^2$ surface even when the exterior spacetime is unchanged.
In fact, for macroscopic black holes (with
$r_5 $ sufficiently large), {\it all} inhomogeneous
Fourier modes become very smooth at the horizon.
Thus, not only are such modes allowed but, for large
$k^2 r_5^2$, they are not restricted by the exterior form of the black
string.  This differs from the case of the homogeneous modes with $k^2 =0$
which must be matched across the horizon.

Can these inhomogeneous modes exist outside the horizon?  
Near $r=0$, $H_5\approx r_5^2 /r^2$ both outside and inside the horizon.
So exterior solutions again satisfy (\ref{inhor}) and have the same 
behavior $r^\alpha$ near the horizon.
However we must now consider the behavior of these modes far away from the
horizon.  While we could work with the exact expressions in terms
of Bessel functions, it is sufficient to note that general
arguments imply that any solution which is sufficiently regular
at the origin must be appropriately singular at infinity\footnote{
This follows from the fact that $\partial_x^2$ is a negative definite
operator on $L^2({\bf R}^4)$ while $k^2H_5$ is positive.}.  Since
the mode with $\alpha = 0$ behaves as a constant near infinity
(in fact, it is just the constant solution), it follows that all
modes with $\alpha >0$ in fact diverge at infinity.  Conversely, 
the modes with $\alpha < -2$ are well behaved near $r = \infty$,
but of course diverge at the horizon.

Hence inhomogeneous modes are not allowed outside of the
horizon, unless they are directly generated by sources in the
exterior region.
For example, if we add a single one brane at $(x_0,y_0)$
by solving 
\begin{equation}
\label{source}
\tilde{\nabla}^2 H_1 = \delta(x - x_0) \delta(y-y_0)
\end{equation}
with appropriate boundary conditions,
then the field for $|x| < |x_0|$ would be composed of modes
with $\alpha >0$ while the field for  $|x| > |x_0|$ would
contain modes with $\alpha <0$.

Inside the horizon, the fact that the inhomogeneous modes diverge at
infinity is not a problem, since this occurs 
in the unphysical region beyond the singularity. Moreover, as discussed
in the previous section, the behavior of these modes in the unphysical
region 
is modified by sources at the singularity. 

\section{Interiors with Waves}

We now consider dynamical solutions to (\ref{main}) where any of the various
fields (except $H_5$) may depend on $u$ as well as the coordinates $x$ and $y$.
Roughly speaking, each of the stationary modes described in the previous
section may be given arbitrary $u$ dependence and so becomes a 
propagating mode.
It is convenient to set
\begin{equation}
\label{wave ansatz}
\partial_u H_1 = \partial_i
A^i 
\end{equation}
so that (\ref{main}) again simply reduces to 
the statement that $H_1$, $K$ and $A_i$
are annihilated by the operator $\tilde{\nabla}^2$ (\ref{deftil}).
The condition (\ref{wave ansatz}) is a natural one if we 
 seek solutions whose sources can be interpreted as (now wiggling)
D-branes.  This is because one can view the source of $A_i$ as the momentum of
the various branes while the source of $H_1$ is the energy
density of the branes.  As a result, (\ref{wave
ansatz}) is like a continuity equation.  In particular, the oscillating
string sources of \cite{cmp} are consistent with this condition.

To illustrate this, let us consider the solution corresponding to 
adding a single D-onebrane oscillating outside the horizon of
the homogeneous black string
background described in section II.
To do so, we
introduce a Green's function $\Lambda(x_1,y_1,x_2,y_2)$
for the elliptic operator $\tilde{\nabla}^2$; that is, $\Lambda$
satisfies
\begin{equation}
\tilde{\nabla}^2 \Lambda(x,y,x_0,y_0) = \delta(x-x_0) \delta(y-y_0)
\end{equation}
and vanishes for large $x^2$.
Then the change in the solution due to adding the onebrane is given by
\begin{eqnarray}
\label{onext}
\Delta H_1(x,y,u) &=& m \Lambda(x,y,x_0,y_0(u)) \cr
\Delta A^i(x,y,u) &=& -m \dot{y}_0^i(u)  \Lambda(x,y,x_0,y_0(u)) \cr
\Delta K(x,y,u) &=& m \dot{y}_0^2(u) \Lambda(x,y,x_0,y_0(u)).
\end{eqnarray}
and describes  a onebrane at $x=x_0$, $y = y_0(u)$.
Such a solution carries the appropriate energy and momentum for 
an oscillating string of tension $m/\kappa^2$ \cite{homa} where
$\kappa^2 = {{(2\pi)^5 g^2 } / {{\sf V} }}$ and $g$
is the asymptotic string coupling.  Note that (\ref{onext}) is an
obvious generalization of the oscillating string solutions of
\cite{cmp}.  
The interpretation of this field as arising from the oscillations of
a onebrane in fact forces the relation (\ref{wave ansatz})
between $A_i$ and
$H_1$.  This is just the statement that the momentum carried by 
a string is determined by its motion.

We can now describe a set of modes in the interior which seem to
be a strong coupling analogue of the D-brane states which
represent the black hole entropy at weak coupling.
To begin, we recall that the integer normalized charges $Q_1$, $Q_5$, and $n$
are related to the parameters $r_1$,
$r_5$, and $p$ by \cite{hms}
\be
\label{charges}
Q_1 = {r_1^2 \ \V \over (2\pi)^4 g}, \qquad Q_5 = {r_5^2 \over g},
\qquad n= {p L^2\ \V \over (2\pi)^6 g^2}
\ee
We will consider the case $Q_5=1$, so that the weak coupling description
of the black hole microstates consists of $Q_1$ onebranes oscillating
inside a single fivebrane. To obtain the strong coupling description
of these states, we will choose sources for (\ref{main}) which
reproduce this behavior.
This argument
cannot, of course, be considered an independent derivation of the counting, 
but does serve to give a definite interpretation of the modes in the 
strong coupling regime. 

To understand the fields generated by an oscillating onebrane, we cannot
use (\ref{onext}) since this applies only to strings oscillating
outside the horizon. It is possible to write down the analog
of (\ref{onext}) for a onebrane oscillating inside the horizon.
We must, of course,
take care to impose the appropriate boundary
condition stated in section III; in particular, in the asymptotic
region beyond the singularity, the solution should approach a constant
with zero total charge. (This insures that the charge seen at the horizon
is given by the sources inside.)
Since the homogeneous mode is fixed by matching to the exterior solution,
we must also insure that the only change in this mode at $r=0$
is the increase in charge and momentum due to the string.
The result is that the change in the solution for
a onebrane oscillating inside the horizon is

\begin{eqnarray}
\label{onint}
\Delta H_1(x,y,u) &=& {m \over {r^2}} - m \Lambda(x,y,x_0,y_0(u)) +
m \Lambda(0,0,x_0,y_0(u)) 
\cr
\Delta A^i(x,y,u) &=& -m \dot{y}_0^i(u)\left[ {1 \over {r^2}} - 
 \Lambda(x,y,x_0,y_0(u))  + 
\Lambda(0,0,x_0,y_0(u)) \right]
 \cr
\Delta K(x,y,u) &=& m\dot{y}_0^2(u)\left[{1 \over {r^2}} - 
\Lambda(x,y,x_0,y_0(u))
+  \Lambda(0,0,x_0,y_0(u)) \right].
\end{eqnarray}
Note that these fields satisfy $\partial_u H_1
= \partial_i A^i$ everywhere.
As discussed earlier, since these sources have positive charge
and are at fixed radius inside the horizon, they must have negative
energy density. This need not concern us, since we will eventually
place them at the singularity.

Since the equations are linear, there is no difficulty considering
$Q_1$ different oscillating strings. To insure that the homogeneous
modes are independent of $u$ (which is needed to keep the horizon regular
\cite{hoya}) we require that their oscillations combine in such 
a way that $\sum_A m_A \dot{y}_A^2(u)$ and
$\sum_A m_A \dot{y}^i(u)$ (where $A$ runs over the various
branes) are in fact independent of $u$.

To obtain the complete solution, we start with $H_5(r) = ({r_5^2 \over r^2} -1)
\theta (r_5 - r)$ which describes a fivebrane at $r=r_5$.
We will assume that the singularity lies at $r=r_5$, i.e., $H_1$ will be
nonzero for all $r< r_5$. This will be true whenever $r_1 >r_5$
(in the absence of oscillations), and the inhomogeneities are small.
For most D-brane states, this is indeed the case\footnote{In particular, if 
one counts only the weakly coupled D-brane states in which the onebranes all
oscillate in nearly the same way, one  still 
reproduces the Bekenstein-Hawking entropy. This equipartition
theorem-like result follows, for example, from section III.B. of
\cite{homa}.}.
We then assume that the source for $H_1$ at the singularity is 
composed of $Q_1$ separate pieces (onebranes)
each carrying a unit quantum of charge.
We assume these onebranes
can oscillate in the $y$ directions (since they are bound to the
five brane at weak coupling) and hence they are described by
four bosonic functions (the $y_0^i(u)$ above) which give the position of 
each onebrane. These sources generate nonzero fields $A_i$ and $K$ 
as well as $H_1$.
In this way, the $4Q_1$ left-moving bosonic modes which account for the black
hole entropy  can be represented by modes living inside the black hole
horizon. Although the sources live at the singularity, the modes extend
out to the horizon. We have described these modes classically, but  since
they satisfy linear equations, it should be straightforward to quantize
them and obtain a direct correspondence between their states and the
quantum states of the D-branes.

One can ask whether these interior waves are restricted by the fact
that there are no analogous waves in the exterior. This question
is studied in detail in appendix A and the answer is no.
As we have seen, these inhomogeneous modes all vanish near the horizon.
The fact that the waves do not cross the horizon 
can be understood
by considering the general metrics (\ref{ansatz}),
with or without waves. Such solutions have a null
Killing field ${{\partial} /{\partial v}}$ whose integral curves
are null geodesics. The waves can be thought of as following these
geodesics, which never cross the horizon.
On the other hand, the homogeneous Fourier mode {\it is} restricted by
the exterior solution. This follows simply from the fact that the area
of the horizon is completely determined by this mode. When the
homogeneous mode is independent of $u$, it is completely determined
by continuity at the horizon.

It is interesting to ask whether the entropy can be counted directly from the
low energy field theory without resorting to the D-brane analysis
to fix the sources at the singularity. (For other approaches to this
question, see \cite{lawi,lawiII,cvts,tsey}.) Naively, there are an infinite
number of modes since the wave number $k$ can be arbitrarily large.
However, since we have been solving the low energy string equations,
it is natural to count only modes with wavelength larger than the string scale.
A heuristic counting of modes (for the case of a single fivebrane)
which yields the right order of magnitude is the following.
There are six fields ($H_1$, $K$, and $A_i$) which are roughly 
 independent components of the solution.   At the order
of magnitude level, we can replace this set by a single field $\phi$ which
satisfies the equation
\begin{equation}
\label{phi}
\tilde{\nabla}^2  \phi = 0.
\end{equation}
Since the $u$ dependence of $\phi$ is unconstrained, the space of solutions
reduces to a single left-moving $1+1$ scalar field for every allowed 
solution to  (\ref{phi}) in the transverse $(x,y)$ space.
Since (for macroscopic black strings) all
solutions allow for quite smooth horizons,
we shall include all solutions below our string scale cutoff.
More precisely, we shall include all solutions for which the wave number
$|k|$ is small enough that the wavelength of the mode on the internal
torus at the horizon is above the string scale in the string metric:
$g_{\mu\nu}^S = e^{\phi/2}g_{\mu\nu}$.
The number of such modes is given by the volume $V_H$ of the
internal four-torus at the horizon in the string metric
in units of the string length,
which is in turn determined (see \ref{ansatz} and \ref{charges}) by the
number of D-branes: $V_H = (2\pi)^4 Q_1/Q_5$.  Since $Q_5=1$ for the case
in question, we find on the order of $Q_1$ left-moving bosonic
field modes.  The exact D-brane counting gives $4Q_1$ such modes,
which agrees at the order of magnitude level.

In the above counting of modes, we assumed that all of the modes
were spherically symmetric.  It is interesting to note, however, that
there are aspherical solutions to the above equations as well.
In fact, such solutions behave much like the ones that we have
already discussed.  For sufficiently high angular momentum, or when
the mode is also inhomogeneous in the internal ($y$) directions, 
such modes are again arbitrarily smooth on the horizon. It is tempting
to try to associate these modes with the $Q_5$ factor in the entropy
when $Q_5 >1$, but it is not yet clear how this will come about.

\section{Discussion}

We have studied the region inside the horizon of an extreme black
hole in string theory.  For the same exterior geometry, we have found
a large number of interior solutions which leave the horizon nonsingular.
These solutions contain modes which propagate entirely inside the
horizon, and can be viewed as generated by sources living on the singularity.
We have seen that one can choose these sources to behave like weakly
coupled D-branes,
and thus obtain a strong coupling description of these
states. In this way,
one avoids the usual conflict over whether the information is localized
near the singularity or the horizon, since  these modes extend from
one to the other.

There are several questions  which remain open.  One of them
is whether it is possible to improve the counting of these modes and precisely
reproduce the black hole entropy directly from the low energy field
theory. To do this, one must understand the role of nonspherically symmetric
modes. In our rough counting in the previous section, we included only
spherically symmetric states despite the
fact that (even for $Q_5 = 1$) the sphere at the horizon is large 
at large coupling,
so that nonspherical modes can be both
smooth and have wavelengths much larger than the string scale
at the horizon.  Another question involves how to extend this
counting to the case with $Q_5 > 1$.  We have focused on
the solutions with RR charges, but it is clear from S-duality that there
are analogous solutions (with the same number of
modes propagating inside the horizon)
carrying NS charges\footnote{We thank A. Tseytlin for this observation.}. 
If the counting of these modes can be shown to 
reproduce the Bekenstein-Hawking entropy, then this would provide an
explanation of the entropy of black holes with NS charges as well.

Although we have discussed only extremal black holes, it is likely
that a near extremal black hole will have similar modes which
may not persist indefinitely, but will be very long lived. 
An object which falls into an
extreme black hole is likely to excite these modes.  What effect
do these interior modes have on Hawking radiation?  Can the information
about what falls in now be recovered in the Hawking evaporation?

To begin to address these questions, one  can consider a single 
oscillating (positively charged) D-string
falling into a black string. There are static solutions with the D-string 
oscillating at any radius outside the horizon  (given by (\ref{onext})),
but not inside. As we discussed
in section II, the static positively charged solutions require a 
negative energy density inside the horizon. If the energy density and
charge are both positive, the string will experience a repulsive force.
In Appendix B we study the motion of an oscillating test string
and find that the string falls smoothly
through the horizon, reaches a minimum radius and expands out into
another asymptotically flat region of spacetime. The state of oscillation
remains completely unchanged. 
To show that  the oscillating string excites the interior modes requires
going beyond the test string approximation. 

Something unusual may happen at the event horizon when a nontest string
approaches. Recall that all nonsingular inhomogeneous modes vanish
at the horizon. The horizon itself always remains homogeneous.
It is not yet clear whether this is just a property of the modes we
have considered (which all preserve a null translational symmetry,
and are likely to be supersymmetric) or whether this property holds
more generally. If it does,
any perturbation outside the horizon
which is inhomogeneous
in the compact directions, must become homogeneous when it crosses the horizon.
It is then likely that the perturbation would  remain homogeneous inside, and
could not excite the
interior modes. However, in this case, it would appear that the
horizon must become singular when a onebrane passes through -- not just
at the particular point $y=y_0$ occupied by the onebrane on the horizon,
but over the entire four-torus.  Perhaps a more plausible alternative 
is that inhomogeneities
are not always smoothed out when an object crosses the horizon, due to
transient modes which do not preserve a null translational symmetry.
In this case, the interior modes that we have discussed can be excited.

\acknowledgements
It is a pleasure to thank A. Strominger and A. Tseytlin
for useful discussions. This
work was supported in part by NSF grant PHY95-07065 and in part
by funds provided by Syracuse University.  Some of this work was
carried out while GH was at the Aspen Center for Physics and
DH was at the Erwin Schr\"odinger Institute. We would like
to thank them for their hospitality and partial support.

\appendix

\section{The Smoothness of the Horizon}

In this appendix we address the question of to what extent our
various modes in the interior of the black string attach smoothly
to the exterior.  Recall \cite{homa} that
the smoothness of the horizon is difficult to analyze when the
homogeneous modes (that is, the $y$ translationally invariant and 
spherically symmetric modes such as $K = p/r^2$) are $u$-dependent.
In fact, it now seems \cite{hoya} that the horizon is actually
singular when such modes are nontrivial functions of $u$.
However, since we are primarily interested in the behavior of
the higher modes, this will not cause a problem; we simply
take the homogeneous modes to be independent of $u$ while allowing
arbitrary $u$ dependence for the higher modes.  Note that
dropping only the lowest modes does not affect the
counting of D-brane states.

For simplicity, we shall study the case
with $r_1 = r_5 = r_0$, although the more general
case may be addressed by the same techniques and yields corresponding
results.   We are interested
in the effect of adding a term $\Delta$ to the field $H_1 = {{r_0}^2
\over {r^2}} -1$ such that $\Delta \sim C_\Delta(u)
r^\delta$ for $\delta>0$ near $r=0$.
We will also add a similar term $k \sim C_k(u) r^{\kappa}$
to $K = {p \over {r^2}}$ and a term $a_i \sim C_{a_i}(u) r^{\alpha_i}$ to
$A_i = {{ r_0^2\dot{f}_i} \over {r^2}}$ ($\kappa,
\alpha_i >0$).  Here, $\dot{f}_i = df_i/du$  but since we keep the
homogeneous modes independent of $u$,
$\ddot{f}_i = 0$ and
$f_i(u) =u  \dot{f}_i$.  We use this notation to
coincide with that of \cite{homa}.  Note that due to the compactness
of the $S^1$ ($z$) direction, the coefficients $C_\Delta(u)$, 
$C_k(u)$, and $C_{a_i}(u)$ are periodic in $u$.  As a result, 
they are bounded but approach no well-defined limit at the horizon
($u \rightarrow -\infty$).

A final simplification will be to conformally transform
the metric by multiplying by 
$e^{-\phi/2}$.  If both $e^\phi$ and the new metric are $C^n$ (and 
$\phi$ is finite), then it follows that the original metric is
$C^n$ as well.  The particular metric to be analyzed is then\footnote{This
is the string metric of the S-dual solution where the RR three form is
exchanged for the NS three form.}

\begin{equation}
ds^2 = H_1^{-1}
du \left( dv + K du
+ 2 A_i dy^i \right)
+ dy^2 + H_5 dx^2;
\end{equation}
that is,
\begin{equation}
ds^2 = \left( {{r_0^2} \over {r^2}} -1 + \Delta \right)^{-1}
du \left( dv + \left( {p \over {r^2}} + k \right) du
+ \left( 2 r_0^2 {{\dot{f}_i} \over {r^2}} + 2 a_i \right) dy^i \right)
+ dy^2 + H_5 dx^2.
\end{equation}
Note that the $A_i = {{ r_0^2 \dot{f}} \over {r^2}}$ term corresponds
to the `internal' waves of \cite{homa}, although our $y$ coordinates
are the $y'$ coordinates of \cite{homa}.  Note also that our
$p$ would be $p - r_0^2 \dot{f}^2$ in the notation of 
\cite{homa,homaII}.

In analogy with the procedure followed in \cite{homa}, we now
introduce new coordinates $ R = r_0 \sqrt{ {r_0^2 - r^2}
\over {r^2}}$, $\hat{y}^i = y^i + f^i$ and $\hat{v} = 
v+ 2 \dot{f}_i \hat{y}^i - \int du \dot{f}^2$
so that the metric becomes

\begin{eqnarray}
ds^2 &=& r_0^2 \Bigl\{  \left( R^2 + r_0^2 \Delta \right)^{-1}
du \Bigl( d\hat{v} + (2a_i - 2 \Delta \dot{f}_i) d\hat{y}
\cr
&+& \left( (p - r_0^2 \dot{f}^2) {{R^2 + r_0^2} \over {r_0^4}} + k 
+ \Delta \dot{f}^2 - 2 a_i \dot{f}^i \right) du \Bigr)
+ r_0^{-2} d\hat{y}^2 + R^{-2} Z^{-3} dR^2 + Z^{-1} d^2 \Omega_3  \Bigr\}.
\end{eqnarray}

A further change of coordinates to
$U =  {1 \over {2 \sigma}} e^{2 \sigma u}$,
$V= \hat{v}
+ 4 r_0^{2} \sigma^2 u - R^2 \sigma$, and $W =
e^{-{\sigma}u} R^{-1}$ (with $\sigma = r_0^{-2} \sqrt{p - r_0^2 \dot{f}^2}$)
places the metric in a form which exactly matches the metric\footnote{Note,
however, 
that we are now in the coordinate range $U >0$ which describes the
interior solution, whereas \cite{homa} worked with the exterior solution
where $U < 0$.  This is just the analytic continuation referred to in
section II.} in
appendix A of \cite{homa} when $k = \Delta = a_i = 0$.
Since the metric $ds_0$ corresponding to $\Delta = k = a_i = 0$
is already known to be smooth, it is sufficient to analyze the deviations
from this metric.  Expanding out the solution we find

\begin{equation}
ds^2 - ds^2_0 = {{r_0^2 W^2} \over {2 \sigma U}} \left( - r_0^2 
\sigma^2 \Delta + k + \Delta \dot{f}^2 - 2 a_i \dot{f}^i \right)
+ higher  \ order  \  in \ U.
\end{equation}

Since $\Delta \sim C_\Delta r^\delta \sim  C_\Delta W^\delta
U^{\delta/2}$, $k \sim C_k r^\kappa
\sim C_k W^\kappa U^{\kappa/2}$,
 and $a_i \sim C_{a_i} r^{\alpha_i} \sim C_{a_i} W^\alpha_i
U^{\alpha_i/2}$ and the coefficients $C_\Delta$, $C_k$, and $C_{a_i}$
are bounded (but not continuous) at $U=0$, this metric is $C^\epsilon$ for any
$\epsilon < {\delta \over 2} -1, { \kappa \over 2} -1, {{\alpha_i} \over 2} -1$.
Similarly, $e^\phi$ is $C^{\gamma/2}$ for all $\gamma < {\delta 
\over 2}$, so the physical
(Einstein) metric is again $C^\epsilon$ for any
$\epsilon < {\delta \over 2} -1 , { \kappa \over 2} -1 , {{a_i} \over 2} -1 $.

\section{Motion of Test strings}

In this appendix we study the motion of an oscillating test
D-string falling into an extremal black hole. We will assume for
simplicity that $r_1=r_5 = r_0$ and that
the black hole itself is not carrying any waves.
Then the black hole metric can be written in the form
(\ref{both})
\be
ds^2 = - F(r) dudv + {p\over r^2} du^2 + F^{-2}(r) dr^2 + r^2 d\Omega_3
+dy_i dy^i
\ee
where
\be
F(r) \equiv 1-\lp{r_0\over r}\rp^2
\ee
and $p$ is constant.
In addition, $B_{uv} = -F$, and there is a nonzero $B_{\mu\nu}$ on the three
sphere which will not play a role in our discussion. The radial coordinate here
(which was denoted $\hat r$ in (\ref{both})) is different from that
used in most of this paper.
The horizon is now at $r=r_0$ and the singularity is at $r=0$.
These coordinates are convenient since they
cover both the regions inside and outside the horizon. Thus we will be
able to follow the motion of the test string across the horizon.

The motion of D-branes is described by a Dirac-Born-Infield action. For
onebranes, this is the same as the usual string action. To begin,
we will assume that there is no motion in the angular directions.
We will include nonzero angular momentum later.
If we use conformal
gauge, and introduce
null coordinates on the worldsheet $\s_{\pm} = \tau \pm \s$,
the sigma model action  takes the form
\be
S \propto \int d\s_+ d\s_- \lp -F\p_+ u \p_- v + {p\over r^2} \p_+ u \p_- u
+ F^{-2} \p_+ r \p_- r + \p_+y_i \p_- y^i\rp
\ee
Notice that the action is invariant under shifting $v$ by an arbitrary
function of $\s_+$. This is a direct result of the null Killing vector
and the fact that $B_{uv}= g_{uv}$ \cite{hots}. In addition to the equations
of motion following from this action, we must satisfy the constraints
\be\label{constp}
-F\pp u \pp v +{p\over r^2}(\pp u)^2 + F^{-2} (\pp r)^2 + (\pp y_i)^2 =0
\ee
\be\label{constm}
-F\pk u \pk v +{p\over r^2}(\pk u)^2 + F^{-2} (\pk r)^2 + (\pk y_i)^2 =0
\ee

The equation of motion for $y^i(\s_+,\s_-)$
is a simple wave equation, $\p^2 y^i
=0$.
Since this equation decouples from the remaining equations of motion,
it follows immediately that the wave on the test string is independent of its
other motion. In particular, if the test string falls into the black string,
it retains the same wave it had outside.
We will assume
that the string carries a right moving wave only: $y^i = y^i(\s_-)$.

The $v$ equation of motion is $\pk(F\pp u) = 0$ which implies that
$F\pp u $ is an arbitrary function of $\s_+$. In conformal gauge, one
has residual gauge freedom to reparameterize $\s_\pm$ separately.
Using this,  one can set $F\pp u $ equal to a constant, which we choose
to write as $E/2$. Thus
\be\label{usol}
\pp u = {E\over 2 F}
\ee
Using this, the constraint (\ref{constp}) becomes
\be\label{vcon}
\pp v = {pE\over 2r^2 F^2 } +  {2(\pp r)^2 \over EF^2}
\ee
The $u$ equation of motion is
\be\label{uequ}
\pp(F\pk v) - \pp\left ( {p\over r^2} \pk u \right )
- \pk \left ( {p\over r^2} \pp u \right ) =0
\ee
We  will
look for a solution where $r$ is a function of $\tau = (\s_+ + \s_-)/2$ only.
Then the
$\pk$ in the last term above can be replaced by $\pp$ (since it acts on a
function of $\tau$) and we can immediately integrate to obtain
\be\label{vsol}
F\pk v -  {p\over r^2} (\pk u + \pp u ) = f(\s_-) +c
\ee
where $f$ is an arbitrary function of $\s_-$ which integrates to zero and $c$
is a constant.

Since $r$ is a function of $\tau$ only, we can relate $\pk u$ to $\pp u$,
and $\pk v$ to $\pp v$ by applying $\p_\s = \pp-\pk$ to
(\ref{usol}) and (\ref{vcon}) to find
$\p_\s \pp u  =0,\ \p_\s \pp v  =0$. So $\p_\s u$ and $\p_\s v$ are both
functions of $\s_-$ only. One might
think that the integral of these functions must vanish. However recall that
our spacetime is identified so that $v-u = 2z$ is periodic, and we
want our test string to wrap around this compact direction.
This implies
\be\label{rel}
\pk u = \pp u + k + g(\s_-) \qquad \pk v = \pp v - k + h(\s_-)
\ee
where both $g$ and $h$ integrate to zero and $k$ is a constant related
to the winding number.

Rather than solve the radial equation directly, we use the constraint
to obtain an energy-like equation. (This is similar to how one describes
the motion of geodesics in spherically symmetric spacetimes.)
Substituting (\ref{rel}) into (\ref{vcon}) and (\ref{vsol}) we obtain
two expressions for $\pk v$. Setting the $\s_-$ dependent terms equal
to each other yields
\be
f=h = - {p\over r^2_0} g
\ee
The remaining terms yield
\be
\dot r^2 + V_0(r) =0
\ee
where
\be\label{pot}
V_0(r) = -E \left[2kF\lp F+{p\over r^2}\right ) +{pE\over r^2} + 2cF\right]
\ee

To complete the solution, we now consider the second constraint
which we take to be the
difference between (\ref{constm}) and (\ref{constp}). Since $r$ is
a function of $\tau$ only, the $(\p r)^2$ terms cancel, and using the above
results we get
\be\label{geq}
g^2 + g\left(k+ {E\over 2} -{c r_0^2 \over p} \right ) +
{r_0^2\over p} \lp {kE\over 2} -ck  + (\pk y^i)^2 \rp =0
\ee
Notice that all $r$ dependent terms have dropped out, and we are left
with a quadratic equation which determines $g(\s_-)$ in terms
of the wave $(\pk y^i)^2$. Since we have assumed that the integral of $g$
vanishes, we determine the arbitary constant $c$ by demanding that this
is the case.

The radial motion of the oscillating test string is
completely determined by the
potential (\ref{pot}). At infinity, $\dot r^2 = 2E(c+k)$,  so $E$ is related
to the initial kinetic energy of the string. The only way that the
oscillations affect the radial motion is through the constant $c$ in
the potential. (The constant $k$ is determined by the winding number.)
In the absence of waves $y^i(\s_-) =0$, the solution to (\ref{geq})
is $g(\s_-)=0$ and $c=E/2$. The potential then becomes
\be
V_0(r) = -E\left[\left(F + {p\over r^2}\right) (2kF +E)\right]
\ee
For positive $E$,
this potential is strictly negative outside the horizon, so the
test string falls into the black hole. However, it doesn't reach
the singularity. The potential vanishes at $r^2 = r^2_0 -p$ and at
$r^2 = 2k r_0^2/(2k+E)$, so the test string turns around at the larger
of these two values. The first corresponds to the location where
$\p/\p z$ becomes null. For small $E$ (or large $p$),
the second value is larger, and the
test string does not penetrate very far inside the horizon.

Since the potential is proportional to $E$, it would appear that 
if $E=0$, every configuration of constant $r$ is a solution. However,
the derivation assumed $E\neq 0$. In light of the comments in section III
concerning the difference between static sources inside and outside
the horizon, it is of interest to study this case more closely. For
simplicity, we will assume there are no waves $y^i(\s_-)=0$. If
$E=0$, then $\p_+ u=0$, so $u=k\s_-$. Trying the solution $v=\alpha \s_+
+ \beta \s_-$, constraint (\ref{constm}) implies $\beta = pk/r^2 F$,
and (\ref{rel}) implies
\be
\alpha = \beta + k = {k\over F} \left(F+ {p\over r^2}\right)
\ee
One can easily check that the remaining equations of motion and constraints
are satisfied. Thus we do
have a solution for a static onebrane at each value of $r$ for which
$F \neq 0$ (that is, away from the horizon).  However, as one might 
expect from the discussion in section III, our static solutions
describe positively charged onebranes outside the horizon and
{\it negatively} charged onebranes inside the horizon.  This follows
from the fact that the orientation on the $(u,v)$ plane is
$du \wedge dv = k\alpha d\s_- \wedge d\s_+$. The sign of the onebrane
charge is thus determined by the sign of $\alpha$.
As long as the onebrane is away from any region of closed timelike
curves, $F + {p \over {r^2}} >0$, so the sign of the onebrane
charge is determined by the sign of $F$, which changes at the horizon.

Nonradial motion can be included just as one does for geodesics. By
spherical symmetry, the test string moves in a plane. Let $\varphi$
denote the angle on the plane, and we will assume that it is only a
function of $\tau$. Then the field equation is $(r^2 \dot \varphi)^. =0$,
which implies $\dot \varphi = L/r^2$ for a constant $L$.
The constraint (\ref{constp})
now picks up an extra term $r^2 (\pp \varphi)^2$ so that the radial
equation becomes $\dot r^2 + V =0$ where
\be
V= V_0 + F^2 {L^2\over r^2}
\ee
and $V_0$ is the potential (\ref{pot}) with no angular momentum.
The angular momentum barrier vanishes near
at the horizon, but is positive outside. So for a given energy,
there is a range of angular momentum
which will still result in capture by the black hole.

\end{document}